\documentclass[prb,aps,twocolumn,amsdtx,showpacs]{revtex4}
\usepackage{graphicx}
\usepackage{amsmath}
\usepackage{amssymb}

\newcommand{\bleq}{\ifpreprintsty
                   \else
                   \end{multicols}\vspace*{-3.5ex}{\tiny
                  \noindent\begin{tabular}[t]{c|}
                  \parbox{0.493\hsize}{~} \\ \hline \end{tabular}}
                   \fi}
\newcommand{\eleq}{\ifpreprintsty
                 \else
                   {\tiny\hspace*{\fill}\begin{tabular}[t]{|c}\hline
                    \parbox{0.49\hsize}{~} \\
                    \end{tabular}}\vspace*{-2.5ex}\begin{multicols}{2}
                    \fi}
\newcommand{\bcols}{\ifpreprintsty\else\begin{multicols}{2}\fi}
\newcommand{\ecols}{\ifpreprintsty\else\end{multicols}\fi}

\newcommand \beq  {\begin{equation}}
\newcommand \eeq  {\end{equation}}
\newcommand \bea {\begin{eqnarray} }
\newcommand \eea {\end{eqnarray}}

\begin{document}
\title{Geometric frustration inherent to the trillium lattice, a sublattice of the B20 structure}
\author{John M. Hopkinson} 
\email{johnhop@physics.utoronto.ca}
\author{Hae-Young Kee}
\email{hykee@physics.utoronto.ca}
\affiliation{60 St.George St., University of Toronto, Toronto, Ontario, Canada}
\date{\today}
\begin{abstract}

We study the classical Heisenberg model on a recently identified three dimensional corner-shared equilateral triangular lattice, a magnetic sublattice to a large class of systems with the symmetry group P2$_1$3.  Since the degree of geometric frustration of the nearest neighbor antiferromagnetic model on this lattice lies on the border between the pyrochlore (not ordered) and hexagonal (ordered) lattices, it is non-trivial to predict its ground state.  Using a classical rotor model, we find an ordered ground state with wavevector $(\frac{2\pi}{3a_0},0,0)$ featuring 120$^o$ rotated spins on each triangle.  However, a mean field approximation on this lattice fails to find an ordered ground state, finding instead a non-trivially degenerate ground state.  As the mean field approach is known to agree with Monte Carlo on the pyrochlore lattice, the reasons for this discrepancy are discussed.  We also discuss the possible relevance of our results to MnSi. 

\end{abstract} 

\maketitle

\section{Introduction}

\begin{figure}
  \includegraphics[scale=0.33]{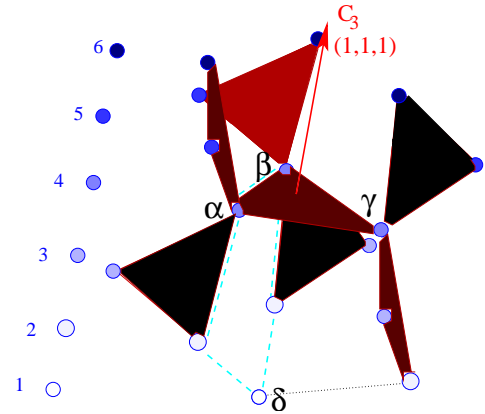}
\caption{\label{figure1} (Color online) The trillium lattice.  Each site belongs to 3 equilateral corner-sharing triangles.  $\{\alpha,\beta,\gamma,\delta\}$ denote the four atoms of the cubic unit cell.  Normal vectors to each triangle point along local (111) axes. (1,1,1) is a C$_3$ rotation axis of the crystal; the front right lower triangle is related to the front left top triangle by translation. In blue (at left) we identify the layer number that each point of the equilateral triangles shown belongs to.  The separation between layers 4 and 5 is 4$ua_0/\sqrt{3}$; between 3 and 4: (1-4$u$)$a_0/\sqrt{3}$ and alternates. After triangles, the next smallest nearest neighbor connection has 5 sides as shown by the dashed line.
}
\end{figure}

Systems with magnetic moments usually order as one approaches zero temperature. However, magnetic systems that form triangle- or tetrahedron-based lattices are often able to avoid ordering until unusually low temperatures, and have been found to have interesting properties both theoretically and experimentally{\cite{Takagi,hermele,mireb,Takada,coldea}}.  Such lattices are generally referred to as geometrically frustrated as, for example, antiferromagnetic interactions between spins render the systems unable to find a unique ground state from magnetic considerations alone.  It has been argued{\cite{moesschalk}} that the degree of frustration of a lattice can be quantified in terms of the connectivity of its triangular units.  This leads to a hierarchy of lattices increasing in frustration from the ordered edge-shared hexagonal lattice, through the corner-shared triangle kagom\'{e}, garnet and $\beta$-Mn lattices, to the corner-shared tetrahedral lattice (common to the pyrochlore, cubic Laves and normal spinel structures).  As we will show, the three dimensional corner-shared equilateral triangle structure of Fig. \ref{figure1} is a fascinating new addition to this hierarchy.  To avoid the cumbersome use of the terminology ``three dimensional corner-shared triangle'' (which also applies to the 'distorted windmill' $\beta$-Mn structure), we will henceforth name this lattice the trillium lattice{\cite{trill}}.  The trillium lattice is a sublattice of the CO (B21), NH$_3$ (D1), NiSSb (F0$_1$, Ullmanite) and FeSi (B20) structures. For example, the Mn atoms of MnSi (B20 structure) form the trillium lattice.    

Recently Pfleiderer et al {\cite{pflei}} reported the observation of a ``partially ordered'' magnetic state in single crystals of MnSi near and above the critical pressure of this itinerant helimagnet. In this neutron scattering measurement, the well-defined magnetic Bragg peak ($q\sim$0.0214 $\AA^{-1}$) with wavevector along (111) was seen to shift to point predominantly along (110) (with $q\sim$0.0304$\AA^{-1}$) upon the application of pressure. The magnetic scattering pattern was seen to feature a sphere at fixed wavevector magnitude, with anisotropic intensities which weakly peaked along (110).  This sphere was found to onset at temperatures consistent with a continuation of the slope of $T_N$ (where helimagnetism sets in). Coincident with the appearance of this unusual magnetic signature an anomalous non-Fermi liquid resistivity $\rho\propto T^{1.5}$ has been observed and found to persist up to twice the critical pressure{\cite{pflei2}}.  The lack of a rapid recovery of $\rho\propto T^2$ in such an apparently clean system ($\rho_0 \sim 0.33\mu\Omega cm$) poses an important challenge to our understanding of Hertz-Millis-Moriya spin-fluctuation theory near a ferromagnetic or helimagnetic{\cite{belitzachim}} quantum critical point.  

Magnetic susceptibility measurements have been found to exhibit high temperature Curie-Weiss tails{\cite{explan}} with Curie-Weiss constants tuning continuously from weakly positive temperatures{\cite{pflei3}}, consistent with the onset of helimagnetic order at ambient pressures, to weakly negative offsets above the critical pressure. This may be indicative of a crossover from ferro- to antiferromagnetically dominated fluctuations, although debate exists in the literature on this point{\cite{thessieu}}. Therefore, we ask whether any link might possibly be made between an enhancement of the antiferromagnetic couplings and the observed quasi-degenerate nature of the spin-spin correlations in this material at high pressures.

   To begin to address this question, we need to understand the role of antiferromagnetic interactions on this lattice. Here we advance a microscopic description of the physics on the magnetic lattice of MnSi.  It is important to notice that the Mn bonds of this lattice form a trillium lattice{\cite{vanderMarel}} (see Fig. \ref{figure1}) whose degree of geometric frustration lies on the border between previously studied frustrated systems and those known to order despite frustration inherent to their lattices.  We first classify the degree of magnetic frustration of the classical Heisenberg antiferromagnet on this lattice, and then use mean field theory and a mapping to a classical rigid rotor model, to explore the corresponding low energy physics.

In our model, we have chosen only to include the localized spin degrees of freedom.  For the majority of magnetic systems forming on this lattice this is clearly an oversimplification, and a more complete model for the monosilicides might be the Kondo lattice model.  Within this model magnetic correlations which arise between neighboring localized spins are mediated by conduction electrons.  The basic magnetic physics of such an RKKY interaction{\cite{RKKY}} is thought to be captured by the extended Heisenberg model.  This paper represents the first step toward such a treatment which will be presented in the near future{\cite{usto}} and is expected to yield a rich phase diagram and potentially account for the small positive Curie-Weiss offset seen in MnSi near its critical pressure.

Mapping the antiferromagnetic Heisenberg model to a classical rotor model, we find an ordered ground state with wavevector ($\frac{2\pi}{a_0}$,0,0) featuring $120^0$ rotated spins on each triangle. Monte Carlo calculations have been carried out{\cite{Sergeius}}  and support this conclusion.  However, the commonly used mean field approximation does not find an ordered ground state.  Rather, mean field theory finds a macroscopically degenerate ground state with degenerate wavevectors forming a sphere-like surface in momentum-space, reminiscent of the partially ordered state of MnSi. Although the detailed anisotropy of the structure factor is not reproduced, to our knowledge this is the first model to find such a partially ordered ground state.  The disagreement between Monte Carlo and mean field approximations is surprising in light of the excellent qualitative agreement between these methods on the kagom{\'e} and pyrochlore lattices, and the reason for this disagreement is discussed.  The strength of magnetic order within the rotor model and its relationship to the mean field results is presented, laying the groundwork for future finite temperature studies where Monte Carlo and mean field results find qualitative agreement to reasonably low temperatures{\cite{Sergeius}}.

The outline of this paper follows.  In Section II an overview of recent excitement in the study of geometrically frustrated materials allows us introduce the trillium lattice in its proper context.  In Section III we employ the standard {\cite{isakov,reimers}} mean field approximation which implicitly relaxes the constraint on the spin ($\sum_{i=1}^4 s_i^2=4$) at each site and calculate the neutron structure factor.
In Section IV we numerically impose a hard spin constraint ($s_i^2=1$) at each site.  We use a classical rotor model to characterize the low-lying excitations and ground state.
  Section V  contains a discussion of the main results of this work. 
\section{Geometric frustration}

\subsection{The trillium lattice}
\begin{table}[hbtp]
\begin{tabular}{|l|l|l|}
\hline
label&position&nearest neighbors\\
\hline
\hline
$\delta$&($u,u,u$) &($\alpha_{-\hat{z}},\alpha_{-\hat{x}-\hat{z}},\gamma_{-\hat{x}},\gamma_{-\hat{x}-\hat{y}},\beta_{-\hat{y}},\beta_{-\hat{y}-\hat{z}}$)\\
$\alpha$&($u$+$\frac{1}{2}$,$\frac{1}{2}$-$u$,1-$u$)&($\delta_{+\hat{z}},\delta_{+\hat{x}+\hat{z}},\gamma,\gamma_{+\hat{z}},\beta,\beta_{-\hat{y}}$) \\
$\gamma$&(1-$u,u$+$\frac{1}{2}$,$\frac{1}{2}$-$u$) &($\delta_{+\hat{x}},\delta_{+\hat{x}+\hat{y}},\alpha,\alpha_{-\hat{z}},\beta, \beta_{+\hat{x}}$) \\
$\beta$&($\frac{1}{2}$-$u,1-u,u$+$\frac{1}{2}$) &($\delta_{+\hat{y}},\delta_{+\hat{y}+\hat{z}},\alpha,\alpha_{+\hat{y}},\gamma,\gamma_{-\hat{x}}$) \\
\hline
\end{tabular}
\caption{\label{Table II} The nearest neighbor sites in relation to the unit cell depicted in Fig. \ref{figure1}.  The subscript denotes translation by one unit cell in the direction noted.} 
\end{table}

The B20 crystal structure to which MnSi belongs contains two interpenetrating sublattices.  For MnSi, one sublattice consists of the magnetic Mn atoms, while the other consists of non-magnetic Si atoms.  Each sublattice forms an infinite three-dimensional lattice of corner-shared equilateral triangles, which we have named the trillium lattice.  The trillium lattice forms a simple cubic lattice with a four site basis.  Elements of this basis are listed in Table \ref{Table II} and shown in Fig. \ref{figure1}.  First, second, and third nearest neighbor bonds are all found{\cite{vanderMarel}} to form networks of independent, corner-shared equilateral triangles, such that even in systems with dominantly ferromagnetic correlations between nearest neighbors, frustration may play a role.  The symmetry properties of this lattice $P2_13(T^4)$ are known to include C$_3$ axes through the center of each triangle and a screw axis, as discussed for example in Bradley and Cracknell{\cite{bradley}}.  

The history of the corner-shared tetrahedral lattice suggests that it is worth investigating systems other than the one of current interest to lay the basis for future work.  It surely was not appreciated by Pauling that his famous estimate{\cite{sergei2}} of the entropy of water ice (a strongly polar molecule) might lead to the explosion of interest in antiferromagnetically correlated Heisenberg spin systems on this lattice.  Harris {\it{et al}}\cite{Harris} found that large spin systems with ferromagnetic correlations could exhibit spin ice behavior with its large low temperature entropy.  Such a realization so many years after Anderson{\cite{anderson}} and Villain's{\cite{villain}} pioneering works on antiferromagnetic spins on this lattice, must have come as a surprise to the early workers on magnetically frustrated lattices.  As Harris {\it{et al}} explained, spin ice physics can arise due to crystal anisotropies which can pin large spins to lie along local Ising directions--symmetries of the crystal structure.  

While the focus of this paper is on the antiferromagnetic Heisenberg model on the trillium lattice, we briefly summarize systems that form the trillium lattice.The trillium lattice is a sublattice of the CO (B21), NH$_3$ (D1), NiSSb (F0$_1$, Ullmanite) and FeSi (B20) structures.  Although both CO and NH$_3$ (D1) are non-magnetic, they have large dipolar electric fields.   Magnetically interesting B20 structure systems include binaries of most of the transition metal elements with Si or Ge, although structures have also been reported{\cite{others}} with Sn, Al, Ga, Hg, Mg and Be as the second member in addition to ternary compounds.  Among these, the ``Kondo insulator''{\cite{gabe}} FeSi has achieved prominence as an important constituent of the earth's mantle.   Fe$_{1-x}$Co$_x$Si  behaves as a doped magnetic semiconductor and exhibits a large anomalous Hall effect which has been argued to have spintronics applications{\cite{Manyala}}, and MnSi is an enhanced mass metal close to criticality.  The FO$_1$ structure is believed to include EuPtSi and EuPdSi, spin $\frac{7}{2}$ systems exhibiting (ferromagnetic) Curie laws from 300K down to 5 K.  M{\"o}ssbauer spectroscopy has shown the europium atoms to be in a divalent (Eu$^{2+}$) 4f$^7$ state{\cite{adroja}}. If crystal fields are such that this large moment aligns along the local (111) axes, this could be a good candidate for the realization of spin-ice physics on the trillium lattice as we will outline in future work{\cite{tbpubl}}. In light of the large number of systems which have shown some evidence for strongly correlated physics and the prospect for future systems to study, it is surprising that to our knowledge, this is the first treatment in terms of local moment based magnetism on the trillium lattice.
\subsection{Itinerant frustrated magnetism}

Upon encountering unusual magnetic signatures on a geometrically frustrated lattice, even for metallic systems, it is common to isolate the magnetic degrees of freedom by first treating a Heisenberg model. A number of frustrated systems show metallic behavior with reasonably enhanced effective masses. Among these, the 12 magnetic sites of $\beta$-Mn (A15 structure) have been shown to form the distorted windmill (corner-shared equilateral triangle) lattice, which has a $P4_132$ symmetry.  The Heisenberg model on the distorted windmill lattice has been 
studied at a mean field level and shown to possess a macroscopic ground state 
degeneracy along the (111) axis{\cite{canlac}}. The magnetic sites of both LiV$_2$O$_4$ (the only d-electron heavy fermion)  and (Y$_{0.97}$Sc$_{0.03}$)Mn$_2$ reside on corner-shared tetrahedral lattices.  As detailed below, this lattice has been the subject of intense theoretical study.   To date all such strongly correlated frustrated metals have shown evidence of strong nearest neighbor antiferromagnetic correlations.

  Theoretically, the classical antiferromagnetic Heisenberg model for O(${\mathcal{N}}$) spins was first solved on the kagom\'{e} lattice where the results were reported to be exact in the ${\mathcal{N}}\rightarrow\infty$ limit{\cite{garanin1}} and recently this treatment was extended to the corner-shared tetrahedral lattice{\cite{garanin2}}. Isakov {\it{et al.}}\cite{isakov} showed that the spin correlations could be understood in terms of a ground state constraint requiring the total spin on each tetrahedron to vanish.   These results agree remarkably well with classical Monte Carlo simulations, in particular reproducing a ``bow-tie''-like structure first found in results of Zinkin et al.{\cite{zinkin}}.  Qualitatively similar structures are seen in quantum treatments of the Heisenberg model on the corner-shared tetrahedral lattice.\cite{canlac1}  Experimental evidence for such quasi-degenerate bow-tie structures has been found in the neutron scattering measurements{\cite{Ballou}} of itinerant (Y$_{0.97}$Sc$_{0.03}$)Mn$_2$.   Moreover, the authors{\cite{isakov}} have pointed out that these results are additionally applicable to both spin ice (which has net ferromagnetic correlations) on the pyrochlore lattice and cubic (water) ice.  In this paper, we study a Heisenberg model on the trillium lattice to understand the geometric frustration inherent to this lattice and its possible relevance to MnSi.

\subsection{Quantifying frustration}
\begin{figure}
\includegraphics[scale=0.6]{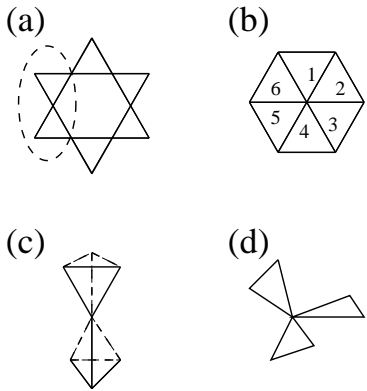}\caption{\label{figure2}The number of units, $b$, common to a given site on common lattices: (a) kagom{\'e}: $b$=2; (b) hexagonal: $b$=6; (c) corner-shared tetrahedra: $b$=2; and (d) corner-shared triangle (trillium, distorted windmill): $b$=3.  }
\end{figure}

For a true energetic minimum, we know{\cite{moesschalk}} that the Heisenberg model can be expressed as,  
\begin{eqnarray}
H&=&J\sum_{\langle ij\rangle}s_i\cdot s_j\nonumber\\&=&J(\sum_{\Delta} |S_{\Delta} |^2 - \sum_i b |s_i|^2)\nonumber\\&=& -b J \hskip1pc \text{per spin}. \label{equation1}
\end{eqnarray}
Here $\sum_{\langle ij\rangle}$ is taken as the sum over all nearest neighbor sites, such that on the lattice each bond is double counted, $S_{\Delta}$ is the total spin on each corner-shared triangular (or tetrahedral) unit of the lattice, and b is the number of connected units at each site as illustrated in Fig. \ref{figure2}.  Here we have used the constraint, $|s_i|^2 =1$.  However, note that the same minimal energy state is reached if one relaxes this constraint to, $|s_i^{\alpha}|^2+|s_i^{\beta}|^2+|s_i^{\gamma}|^2+|s_i^{\delta}|^2 = 1$ where $\alpha$, $\beta$, $\gamma$ and $\delta$ might represent different sites in the unit cell. Minima (for $J>0$) occur when $|S_{\Delta}|^2$ = 0 and have energy -$bJ$ (per spin).   Note the difference between the minimal energy state on the trillium lattice (-3$J$) and that of the corner-shared tetrahedral lattice (-2$J$).  

It is instructive to classify the degree of frustration of the trillium lattice in relation to previously studied lattices (see Table \ref{Table I}).
Counting of the number of degrees of freedom minus the constraints to maintain the ground-state condition {\cite{moesschalk}}, $|S_{\Delta} |^2=0$, we find, the number of degrees of freedom of the ground state, $D$, to be given by,
\begin{equation}
\frac{D}{\bar{N}}=\frac{\tilde{q}({\mathcal{N}}-1)}{b}-{\mathcal{N}}\hskip1pc \label{equation2}
\end{equation}
for $|s_i|^2 =1$, and
\begin{equation}
\frac{D}{\bar{N}}=\frac{\tilde{q}({\mathcal{N}}-\frac{1}{4})}{b}-{\mathcal{N}}\label{equation3}
\end{equation}
for $|s_i^{\alpha}|^2+|s_i^{\beta}|^2+|s_i^{\gamma}|^2+|s_i^{\delta}|^2 = 4$.
\begin{table}[hbtp]
\begin{tabular}{|l|l|l|l|l|l|}
\hline
system&$\frac{D}{\bar{N}}$&$\tilde{q}$&${\mathcal{N}}$&$b$&ground state\\
\hline
corner-shared&1&4&3&2&degenerate\\
 tetrahedra&&&&&\\
\hline
kagome&0&3&3&2&degenerate{\cite{elser,moesschalk}}\\
\hline
garnet&0&3&3&2&degenerate{\cite{taras}}\\
\hline
distorted windmill&-1&3&3&3&degenerate at mf{\cite{canlac}}\\
\hline
trillium  &-1&3&3&3&this work\\
\hline
hexagonal&-2&3&3&6&non-degenerate\\
\hline
\end{tabular}
\caption{\label{Table I} Degeneracy of the ground state of the classical Heisenberg model for several common triangle/tetrahedra based lattices. From Eq. \ref{equation2} and Eq. \ref{equation3}, a na{\"i}ve count of the number of degrees of freedom available to the ground state helps to classify the degree of frustration of a lattice.  For the trillium lattice, the hard spin constraint (rotor model) yields $D/\bar{N}= -1$, which is found to order; the soft spin constraint (mean field) leads to $D/\bar{N}=-1/4$ which seems to exhibit a partial ordering.  Note that the Heisenberg model is believed to order by disorder on the kagom{\'e} lattice\cite{chalker,reimersII} so strictly speaking having a macroscopic number of zero energy spin structures is not sufficient to avoid order.  The spin-ice model on the trillium lattice will necessarily break the $|S_{\Delta}|^2=0$ condition, so Eq. \ref{equation2} and Eq. \ref{equation3} do not apply to this model and a larger ground state degeneracy is possible. 
}
\end{table}
Here $\tilde{q}$ is the number of spins per triangular/tetrahedral unit,  $\bar{N}$ is the number of units, and ${\mathcal{N}}$ is the number of spin components (${\mathcal{N}}$=1 Ising and ${\mathcal{N}}$=3 Heisenberg).
We see that upon strongly enforcing the constraint in Eq. \ref{equation2} (rotor, Monte Carlo) the Heisenberg model is expected to order.   The mean field approximation enforces the relaxed constraint in Eq. \ref{equation3} to produce a less negative value of $D$.    From now on we will refer to the constraint of Eq. \ref{equation2} as the hard spin constraint, while that of Eq. \ref{equation3} will be referred to as a soft spin constraint.  

From this na{\"i}ve counting argument, we would expect to find a degenerate ground state only when $D>0$.  However, it is known that this counting argument can go wrong if the constraints on the spins are not independent{\cite{moesschalk}}, such that a macroscopic degeneracy of the ground state can be present as seen in Table {\ref{Table I}}.  Note that despite this ground state degeneracy, the classical Heisenberg model on the kagom{\'e} lattice is believed to order by disorder in the $T\rightarrow 0$ limit.{\cite{chalker,reimersII}}



\section{A mean field approximation}


\subsection{A remarkable degeneracy}

The nearest neighbor Heisenberg model can be expressed as, 
\begin{equation}
H=J\sum_{\langle ij\rangle}s_i\cdot s_j=\frac{2J}{N}\sum_qS_qM_q^{ab}S_{-q},\label{equation4}
\end{equation}
Here $J>$0 corresponds to antiferromagnetic interactions, $N$ is the number of unit cells in the lattice and $S_q=\left(\begin{array}{cccc}s_q^{\delta},&s_q^{\alpha},&s_q^{\gamma},&s_q^{\beta}\end{array}\right)$.  The four sites of the unit cell are labelled by $\{\delta,\alpha,\gamma,\beta\}$ as shown in Fig. {\ref{figure1}} and Table {\ref{Table II}}.  The second equation has been obtained by the Fourier transformation, $s_i^{\kappa}=\frac{1}{N}\sum_q s_{q}^{\kappa} e^{iqr_{i}^{(\kappa)}}$, where $\kappa$ is chosen from $\{\delta,\alpha,\gamma,\beta\}$ and r$_i^{(\kappa)}$ demarks both the unit cell and the position within the unit cell, with,

\begin{widetext}
\begin{eqnarray}
M_{q}^{ab}=\left(\begin{array}{cccc}0&\cos(\frac{q_x}{2})e^{i((2u-\frac{1}{2})q_y + 2u q_z)}&\cos(\frac{q_y}{2})e^{i(2u q_x+(2u-\frac{1}{2})q_z )}&\cos(\frac{q_z}{2})e^{i((2u-\frac{1}{2})q_x+2u q_y)}\\\cos(\frac{q_x}{2})e^{-i((2u-\frac{1}{2})q_y + 2u q_z)}&0&\cos(\frac{q_z}{2})e^{i((2u-\frac{1}{2})q_x - 2u q_y)}&\cos(\frac{q_y}{2})e^{i(2u q_x-(2u-\frac{1}{2})q_z )}\\\cos(\frac{q_y}{2})e^{-i(2u q_x+(2u-\frac{1}{2})q_z )}&\cos(\frac{q_z}{2})e^{-i((2u-\frac{1}{2})q_x - 2u q_y)}&0&\cos(\frac{q_x}{2})e^{i((2u-\frac{1}{2})q_y - 2u q_z)}\\\cos(\frac{q_z}{2})e^{-i((2u-\frac{1}{2})q_x+2u q_y)}&\cos(\frac{q_y}{2})e^{-i(2u q_x-(2u-\frac{1}{2})q_z )}&\cos(\frac{q_x}{2})e^{-i((2u-\frac{1}{2})q_y - 2u q_z)}&0\end{array}\right),\label{equation5}
\end{eqnarray}
\end{widetext}
where $u=0.138$ for MnSi.
\begin{figure}
\includegraphics[scale=0.2]{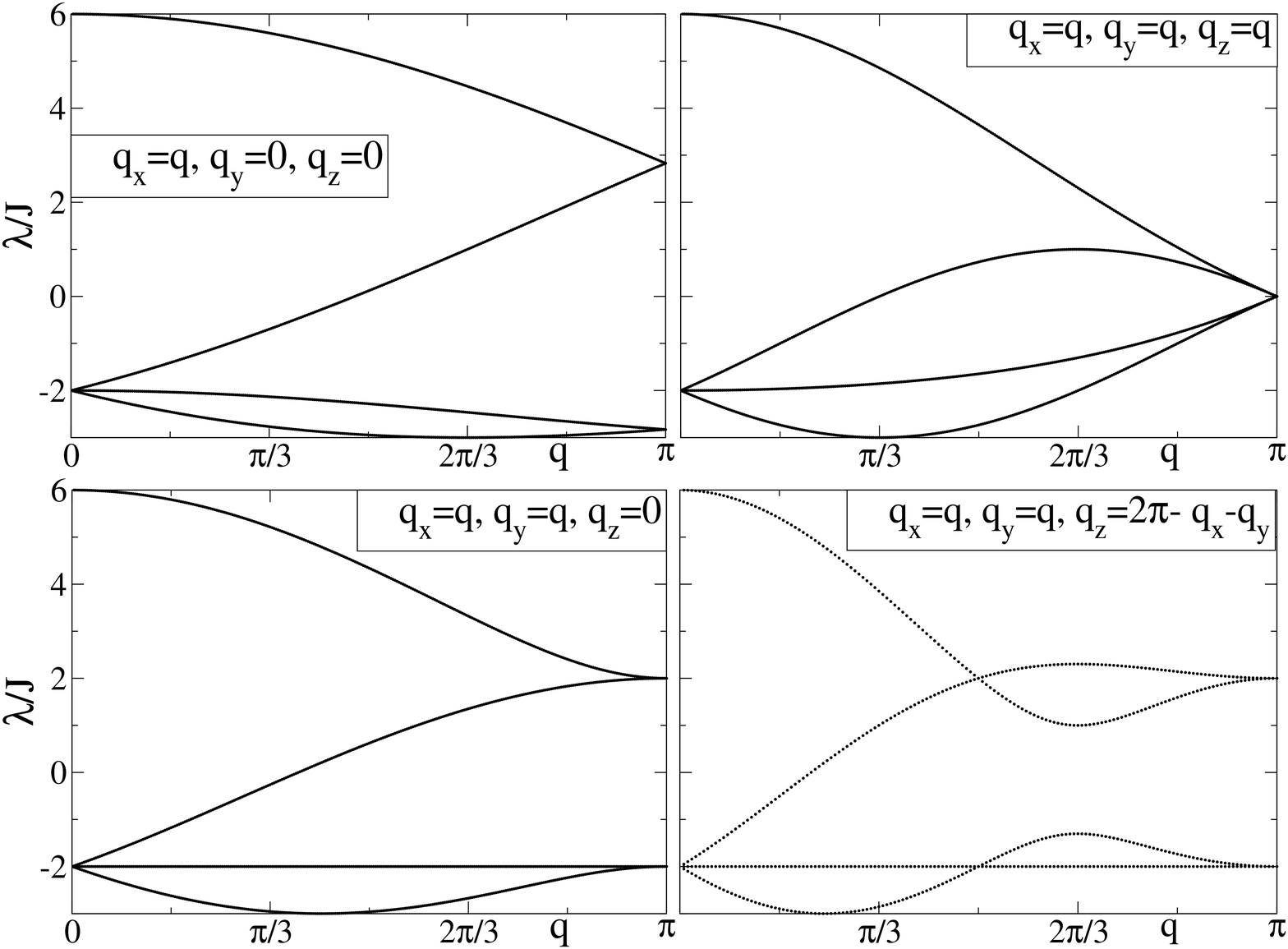}
\caption{\label{figure4} Dispersion curves for $J>$0, $a_0=1$. Notice the degenerate minimum at energy -3$J$, and in the lowest graph a band crossing. The total energy spectrum is invariant under independent operations of $q_x\rightarrow -q_x$, $q_y\rightarrow -q_y$ and $q_z\rightarrow -q_z$.    
}
\end{figure}

Within the mean field approximation, the energy dispersion curves, $\lambda^a(\overrightarrow{q})$ can be obtained by diagonalization of the matrix $M_q$.  Here $a=\{1,..,4\}$ labels the four bands arising from the 4-site unit cell.  In Fig. {\ref{figure4}}, energy dispersion curves are plotted for various directions of $\overrightarrow{q}$.  The ground state is given by the minimum of $\lambda^a(\overrightarrow{q})$.    
A remarkable feature of this spectrum is that the minimum energy solution is degenerate, $u$-independent, and occurs on a sphere-like surface (see Fig. \ref{figure5}) demarked by,
\begin{equation}
\cos^2(\frac{q_xa_0}{2})+\cos^2(\frac{q_ya_0}{2})+\cos^2(\frac{q_za_0}{2})=\frac{9}{4},\label{equation6}
\end{equation}
where we have restored $a_0$, the length of the unit cell for clarity.  Hence, within this approximation, at $T$=0, the classical solution of the nearest neighbor Heisenberg model on the trillium lattice has a degenerate (or partially ordered) ground state.  This apparent degeneracy is a hallmark of frustrated magnetism.  

Note that, for each direction of $\overrightarrow{q}$, a minimum of the dispersion occurs at a single wavevector magnitude, but there are infinite directions of $\overrightarrow{q}$ satisfying Eq. \ref{equation6} with the same minimal energy state, $-3J$.  The partial order of the classical ground state observed here (with a finite $\overrightarrow{q}$ minimum) is qualitatively different from the completely degenerate ground state observed in the corner-shared tetrahedral, distorted windmill and kagom{\'e} lattices, whose dispersions show at least one flat band for their ground state{\cite{reimers,canlac}}.   This implies that the trillium lattice is less frustrated than both the corner-shared tetrahedral lattice and the kagom{\'e} lattice.  This was expected from the na{\"i}ve counting of Table \ref{Table I} as three triangles must meet at a point vs. two for the kagom{\'e} lattice. However, the different behavior of the distorted windmill lattice was quite surprising to us. The trillium and distorted windmill lattices have remarkably different dispersions given the qualitative similarity of their local structures.

Additionally, it is curious to note that along certain directions\cite{cyclic} in momentum space there is a dispersionless band at finite energies.  Peculiarly, a band crossing occurs where this can become the lowest energy state for a given wavevector.  If a similar band structure arises within a strongly correlated nearest neighbor hopping model, extensions of this work might favor the creation of heavy mass particles at chemical potentials close to this energy.  Such a hybridization model is common to most heavy fermion descriptions and the flat bands of the corner-shared tetrahedral lattice have been argued to lead to an explanation for the unusual physics of the only d-electron heavy fermion, LiV$_2$O$_4${\cite{me}}.

\subsection{Spin-spin correlation functions}

To calculate the spin-spin correlation functions and neutron scattering structure factor within this approximation, it is helpful to write the partition function.  Following Isakov et al{\cite{isakov}}, we introduce {$\mathcal{N}$} spin components and constrain each site to have unit spins, 
\begin{eqnarray}
Z&=&\int{\mathcal{D}}S{\mathcal{D}}\lambda e^{-\sum_{\langle ij\rangle}\sum_{\nu=1}^{\mathcal{N}}(\beta Js_i^{\nu}\cdot s_{j}^{\nu} + i\lambda_i \delta_{i,j}(s_i^{\nu}s_i^{\nu}-1))} \label{equation7}\\
&=&\int{\mathcal{D}}S{\mathcal{D}}\lambda e^{-\frac{1}{N}\sum_q \sum_{\nu}S_{q}^{\nu}(\beta(2JM_q^{ab}-\mu 1) + i\lambda)S_{-q}^{\nu}}\\\nonumber&&\times e^{i\lambda{\mathcal{N}}4N}.\label{equation8}
\end{eqnarray}

Note that in moving from Eq. \ref{equation7} to Eq. \ref{equation8} we have implicitly made the mean-field assumption $\lambda_i = \lambda$ for the Lagrange multiplier.  This is in fact an assumption inherent to the treatment of the corner-shared tetrahedral lattice by Isakov et al{\cite{isakov}}, which was found to agree remarkably well with Monte Carlo results{\cite{zinkin}}.  As the four sites within the unit cell are in principle different\cite{equivalent}, this is a four-fold relaxation of the constraints on the system, and implies that we are only  enforcing the single-spin constraint on average within the unit cell.  Additionally, we have introduced a chemical potential $\mu$ in such a way as to shift the value of the minimal energy to 0, and defined $S_q^{\nu}=\left(\begin{array}{cccc}s_q^{\delta,\nu},&s_q^{\alpha,\nu},&s_q^{\gamma,\nu},&s_q^{\beta,\nu}\end{array}\right)$.  To find the saddle-point solutions, we need to minimize Eq. \ref{equation8} with respect to $\lambda$.  Introducing ${\mathcal{S}}_q^{\nu}=S_{q}^{\nu}U$ and $U^{-1}S_{-q}^{\nu}={\mathcal{S}}_{-q}^{\nu}$ to diagonalize $\tilde{M}_q^{aa}=U^{-1}M_q^{ab}U-\mu 1$, we obtain,
\begin{eqnarray}
Z&=&\int\prod_{q,\nu}{\mathcal{D}}{\mathcal{S}}_{q}^{\nu}{\mathcal{D}}\lambda e^{-\frac{\beta}{N}\sum_{q,a,\nu}{\mathcal{S}}_q^{\nu}(2J\tilde{M}_q^{aa} + i\frac{\lambda}{\beta}){\mathcal{S}}_{-q}^{\nu}}e^{i 4\lambda {\mathcal{N}}N}\nonumber\\&=&\int{\mathcal{D}}\lambda\frac{e^{i4\lambda N {\mathcal{N}}}}{\prod_{q,a} \det(\frac{\beta}{N}(J\epsilon_q^a+i\frac{\lambda}{\beta}))^{\mathcal{N}}}\nonumber\\&=&\int{\mathcal{D}}\lambda e^{i4\lambda{\mathcal{N}}N} e^{-\sum_{q,a}{\mathcal{N}}\ln(\frac{\beta}{N}(J \epsilon_q^{a} + i\frac{\lambda}{\beta}))}\label{equation9}
\end{eqnarray}
 where $\epsilon_q^{a}=\lambda_q^a +3 J$ is the energy of the a-th band.  The constraint on $\lambda$ is found to be, $4N=\sum_{q,a}\frac{1}{ \beta J \epsilon_q^{a} +i\lambda}$.  To calculate the spin-spin correlators, we introduce an auxiliary field ${\mathcal{\rho}}_{q}^{\nu}=\left(\begin{array}{cccc}\rho_q^{\delta,\nu},&\rho_q^{\alpha,\nu},&\rho_q^{\gamma,\nu},&\rho_q^{\beta,\nu}\end{array}\right)$ coupling to the spins as,
\begin{widetext}
\begin{eqnarray}
&& Exp\left[{-\frac{1}{N}\sum_q\sum_{\nu=1}^{{\mathcal{N}}}\left(\left(\begin{array}{cccc}s_q^{\delta,\nu},&s_q^{\alpha,\nu},&s_q^{\gamma,\nu},&s_q^{\beta,\nu}\end{array}\right)\left(\begin{array}{c}\rho_{-q}^{\delta,\nu}\\\rho_{-q}^{\alpha,\nu}\\\rho_{-q}^{\gamma,\nu}\\\rho_{-q}^{\beta,\nu}\end{array}\right) + \left(\begin{array}{cccc}\rho_{q}^{\delta,\nu}&\rho_{q}^{\alpha,\nu}&\rho_{q}^{\gamma,\nu}&\rho_{q}^{\beta,\nu}\end{array}\right)\left(\begin{array}{c}s_{-q}^{\delta,\nu}\\s_{-q}^{\alpha,\nu}\\s_{-q}^{\gamma,\nu}\\s_{-q}^{\beta,\nu}\end{array}\right)\right)}\right]\nonumber\\&=&
e^{-\frac{1}{N}\sum_q\sum_{\nu=1}^{\mathcal{N}} ({\mathcal{S}}_q^{\nu}U^{-1}{\mathcal{\rho}}_{-q}^{\nu} + {\mathcal{\rho}}_q^{\nu} U {\mathcal{S}}_{-q}^{\nu})}\label{equation10}
\end{eqnarray}
\end{widetext}
 and calculate $\langle s_{\kappa,q}^{\nu}s_{\kappa',q}^{\nu'}\rangle=\frac{1}{Z}\frac{\partial^2Z}{\partial \rho_{\kappa,-q}^{\nu}\partial \rho_{\kappa',q}^{\nu'}}|_{\rho = 0}$.  After some algebra and a Hubbard-Stratonovich transformation, one arrives at,
\begin{eqnarray}
Z=\int{\mathcal{D}}\lambda \frac{e^{i4\lambda{\mathcal{N}}N}e^{\sum_{q,\nu,a}\tilde{\rho}_{a,q}^{\nu}\frac{1}{J\beta\epsilon_q^a + i\lambda}\tilde{\rho}_{a,-q}^{\nu}}}{(\prod_{a,q}(J\epsilon_q^a + \frac{i\lambda}{\beta})\frac{\beta}{2N})^{\mathcal{N}}},\label{equation11}
\end{eqnarray}
where $\tilde{\rho}_{a,q}^{\nu} = \sum_{l=1}^4\rho_{l,q}^{\nu}U_{la}$ and $\tilde{\rho}_{a,-q}^{\nu} = \sum_{l'=1}^4U_{al'}^{-1}\rho_{l',-q}^{\nu}$.  The saddle-point solution is then
\begin{eqnarray}
\langle s_{\kappa,q}^{\nu}s_{\kappa,-q}^{\nu'}\rangle&=&\delta^{\nu\nu'}\sum_{a=1}^4\{\frac{U_{\kappa' a}U_{a\kappa}^{-1}+U_{\kappa a}U_{a\kappa'}^{-1}}{2(\beta J\epsilon_q^a + i\lambda)}\}.\label{equation12}
\end{eqnarray}
As $U$ is generally not real, we note that $U_{ai}^{-1}=U_{ia}^*$. 


\subsection{The static structure factor}
\begin{figure}
\includegraphics[scale=0.6]{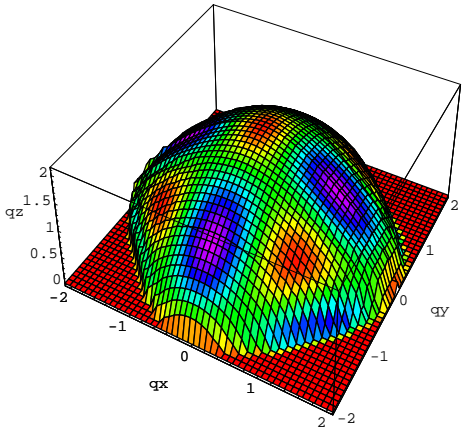}
\caption{\label{figure5} (Color online) Within the mean field approximation, the ground state is found to be degenerate around the sphere-like surfaces corresponding to solutions of Eq. \ref{equation6}.  Here we show (the $q_z=0$ half of) the degenerate spheroid centered at ${\bf{q}}=(0,0,0)$. When $T\rightarrow$0$^+$ and $J>$0, as shown in Fig. \ref{figure6}, one expects at low temperatures a surface of magnetic Bragg peaks to arise in the neutron scattering structure factor coincident with the degenerate spheroids.  Within each Brillouin zone, the relative weight of each of the Bragg peaks is given by the geometrical factor of Eq. \ref{equation14}.  Interference between the two length scales (cubic unit cell, nearest neighbor distance) leads to unusual patternings of intensity of the static structure factor, $S({\bf{q}})$, around each degenerate spheroid. The pattern covering this surface in the first Brillouin zone shows areas of high intensity (purple/dark) along (110), moderate (yellow) around the surface and vanishes along (100) and (111) (red).  However, the overall magnitude of the signal is approximately 1000 times smaller than that seen in MnSi, and this pattern does not hold about the ($\frac{2\pi}{a_0}$,$\frac{2\pi}{a_0}$,0) lattice Bragg point (although the signal would be substantial there).
}
\end{figure}
\begin{figure}
\includegraphics[scale=0.33]{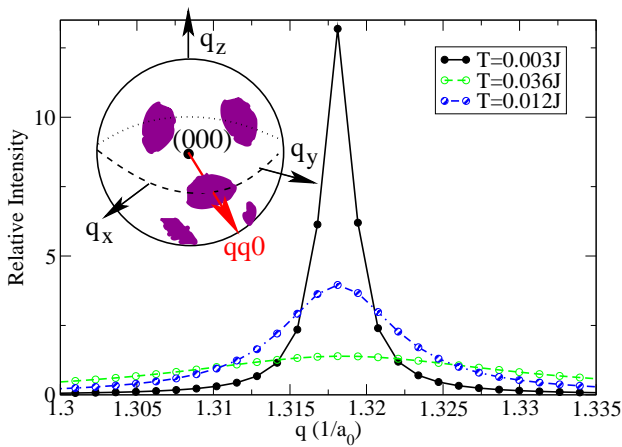}
\caption{\label{figure6} (Color online) Along the $qq0$ direction, the temperature dependence of the longitudinal structure factor.  (Note that unlike in MnSi, the area under the curves seems to be approximately conserved for us). (inset) The structure factor is calculated along the longitudinal direction parallel to $qq0$ as shown.     
}
\end{figure}

The static structure factor can now be found as{\cite{garanin1}},
\begin{equation}
S(\overrightarrow{q})\propto\sum_{{\bf{r}},{\bf{r'}}}\langle s_{{\bf{r}}}^{\perp}s_{{\bf{r'}}}^{\perp} e^{i{\bf{q}}\cdot({\bf{r-r'}})}\rangle\propto\sum_{\kappa,\kappa'}\langle s_{\kappa',q}s_{\kappa,-q}\rangle,\label{equation13}
\end{equation}
under the assumption that for classical spins all spin components contribute equally.

As $T\rightarrow$0$^+$, the small $|\overrightarrow{q}|$ structure factor of the nearest neighbor Heisenberg model has all of its weight on the quasi-spheres given by Eq. \ref{equation6}.  However, the geometry of the lattice leads to an unequal distribution of this weight as a function of angle, and (see Fig. \ref{figure5}) a disperse angle-dependence. The maximal intensity of the structure factor on this spheroid is found along (110) and vanishes due to symmetries of the lattice near (111) and (100).  However, there is actually very little weight within the entire first reciprocal lattice zone of this non Bravais lattice, as for the corner shared tetrahedral lattice.  The sphere-like shape (not shown) centered about ($\frac{2\pi}{a_0}$,0,0) has a distinctly different weight distribution, tuning from minimal weight near ($\frac{4\pi}{3a_0}$,0,0) to maximal near ($\frac{8\pi}{3a_0}$,0,0) with the absolute magnitude of the weight 1000-fold larger yet qualitatively similar temperature dependence.   Translation by a reciprocal lattice vector of the cubic unit cell finds another spheroid, with varying weight and distributions.  One is reminded of the disperse neutron scattering patterns seen on the corner-shared tetrahedral lattice{\cite{garanin2,isakov}}.  There, ``bow-tie'' structures appear as a result of interference between the two length scales given by the unit cell and the structure within the unit cell.  On the trillium lattice one again has two length scales which leads to interesting interference patterns, but at low temperatures, energetic concerns restrict the neutron scattering weight to lie only at wavevectors given by Eq. \ref{equation6}.  The convolution of this energetic structure with the underlying interference pattern produces the varying weights of the structure factor on the sphere-like shapes around different reciprocal lattice vectors.

As in the kagom{\'e} and corner-shared tetrahedral lattices, our model (with relaxed constraint) has a macroscopic ground state degeneracy.  To calculate the structure factor, as shown in Eq. \ref{equation12} and \ref{equation13}, we do a simple sum over all classical spin configuration states weighted according to their energies. 
As the energy gap along the surface of the sphere-like shape is zero, at sufficiently low but finite tempertures, the system will predominantly sample the different ground state spin configurations.  In Fig. \ref{figure5} we plot the maximal value of $S(q)$ as $T\rightarrow 0^+$ for each wavevector direction within the first Brillouin zone to illustrate the idea of degenerate spheroids.  

For the trillium lattice, the second and higher bands are reasonably well-separated from the lowest energy state.  At sufficiently low temperatures compared to $J$,  the lowest band dominates the accessible spin structures.  One expects a strong favoring of a particular wavevector magnitude for each direction in momentum space.  This is seen as a function of temperature along (110) in Fig. {\ref{figure6}}.  

It is instructive to investigate the properties of the extreme limit as T$\rightarrow 0^+$ when only the ground-state spin configurations should contribute appreciable weight to the correlation functions.  In this limit, analytic calculations can be done{\cite{isakov}}, as the structure factor becomes proportional to,
\begin{eqnarray}
S(\overrightarrow{q})\propto\sum_{i=1}^4 U_{i1} U_{1i}^{-1}+\sum_{i\ne j=1}^4 U_{i1} U_{1j}^{-1}\label{equation14}
\end{eqnarray}   
We see that the first term returns 1 as long as the spin eigenfunctions are normalized.  The rest are expressible in terms of the relative spins on each site in momentum space.  For example, a $q$=0 ferromagnetic state would have simply $U_{i1}=U_{1i}^{-1}=\frac{1}{2}$, leading to a factor of 4 in total.  One can show in this limit that Eq. \ref{equation14} vanishes everywhere along (111) by symmetry, although local (111) states about the reciprocal lattice point $(\frac{2\pi}{3a_0},\frac{2\pi}{3a_0},0)$ do not vanish.  The detailed nature of the structure factor pattern on the degenerate spheroid centered at $(\frac{2\pi}{3a_0},\frac{2\pi}{3a_0},0)$ does not correspond to that seen in MnSi.

\subsection{``Spin'' structures: the soft spin constraint}

It is useful to have a physical picture of how the classical spins are arranged along the lattice in the degenerate minimal energy ground state.  From studies on the corner-shared tetrahedral lattice, we are used to the constraint that all spins on each tetrahedron must add to a spin-0 state to be included in the degenerate ground state manifold.  The analogous condition for the trillium lattice (as presented in Sect. II C) is for the spins on each triangle to add to zero.  If we have a truly degenerate ground state then at any finite temperature the classical spins will explore all these states.  

The eigenvectors of Eq. \ref{equation4} hold information about the relative angles between spins.  At any point in momentum space, if the eigenvector is non-degenerate then only one relative angle between spins is contained in this formalism.  Therefore, if a spin structure satisfying the hard spin constraint can be formed in the ground state, its spins will necessarily be coplanar.  However, if such a spin structure cannot be formed in the ground state, then the degeneracy will be seen as an artifact of the soft constraint approximation.

To obtain a particular spin configuration of the ground state, we need to perform a restricted Fourier transform, $s_i^{\kappa}=\frac{1}{N}\sum_{\overrightarrow{q}}e^{i\overrightarrow{q}\cdot \overrightarrow{r}_i^{\kappa}} s_{\overrightarrow{q}}^{\kappa}$.   The relative position vectors are given by, $r_i^{\alpha}=r_i^{\delta} + (\frac{1}{2},\frac{1}{2}-2u,1-2u)$, $r_i^{\gamma}=r_i^{\delta} + (1-2u,\frac{1}{2},\frac{1}{2}-2u)$, and $r_i^{\beta}=r_i^{\delta} + (\frac{1}{2}-2u,1-2u,\frac{1}{2})$.  For simplicity, we consider ``spin'' structures corresponding to the first Brillouin zone in just three high symmetry directions, those with $\hat{q}$ along $(111)$, $(110)$ and $(100)$.   
\subsubsection{(111)}

\begin{figure}
\includegraphics[scale=0.28]{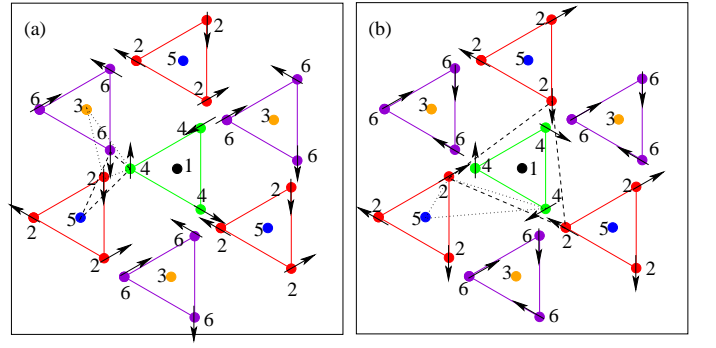}
\caption{\label{figure7} (Color online) Looking down the (111) axis, ``spin'' configurations with the relaxed constraint of Sect. III A. Numbers denote the layer number (or height) within the structure.  (a) The lowest energy $q_0=\frac{\pi}{3a_0}$ state along (111) is shown here.    Dashed and dotted lines label triangles out of the plane.  On these triangles the ground state has two antiparallel spins. (b) A ``spin'' configuration at higher energy along $(111)$ at the same wavevector, which coincidentally would be in the ground state manifold for either second or third nearest neighbor models.  Note that one no longer has antiparallel spins on the small triangles of (a) out of the plane, but that on the dotted lines in (b) (corresponding to 3rd nearest neighbor bonds) one does have antiparallel spins out of the plane.     
}
\end{figure}

To draw the corresponding ``spin'' structure with $\hat{q}$ along (111) at $q_0=\frac{\pi}{3a_0}$, we take the hermitian conjugate of the right ground state eigenvector of Eq. \ref{equation5} (which corresponds to ${\mathcal{S}}_{-q-q-q}$) to obtain, ${\mathcal{S}}_{\overrightarrow{q}}=(0,\frac{1}{\sqrt{3}},\frac{e^{\frac{i2\pi}{3}}}{\sqrt{3}},\frac{e^{\frac{i4\pi}{3}}}{\sqrt{3}})\delta(\overrightarrow{q}-\overrightarrow{q_0})$ where $\overrightarrow{q_0}=(\frac{\pi}{3a_0},\frac{\pi}{3a_0},\frac{\pi}{3a_0})$, which yields,$\{s_{i}^{\delta},s_{i}^{\alpha},s_{i}^{\gamma},s_{i}^{\beta}\}$=$\frac{1}{\sqrt{3}} e^{i(\frac{\pi}{3},\frac{\pi}{3},\frac{\pi}{3})\cdot r_i^{\delta}+\frac{i2\pi(1-2{u})}{3}}$ $\{0,1,e^{\frac{i2\pi}{3}}, e^{\frac{i4\pi}{3}}\}$. The ``spin'' angle gains an additional factor of $e^{\frac{i\pi}{3}}$ by the translation of one lattice constant along any of the $\{\hat{x},\hat{y},\hat{z}\}$ directions of the cubic unit cell.  By inspection, we see that spin contributions on the $\alpha,\beta$ and $\gamma$ sites are of equal magnitude, but are rotated by 120$^0$ relative to one another.  A pictorial representation of this state is shown in Fig. \ref{figure7}(a).  

In this structure every triangle is found to have total spin 0, as expected from Eq. \ref{equation1}.  However, the $\delta$ site (labelled 1,3,5 in Fig. \ref{figure7} (a)) has spin 0.  Triangles involving this site have their other two spins arranged in an antiparallel fashion.  As discussed in Section IV D, unlike on the corner-shared tetrahedral lattice, on the trillium lattice one cannot add perpendicular spin components with a different $\overrightarrow{q}$ to restore the hard spin constraint while remaining in the ground state.     
\subsubsection{(110)}

Within the mean field approximation, the largest neutron scattering structure factor in the first Brillouin zone was found to lie along (110).  The minimal energy spin structure is found at $q_0=\frac{2}{a_0}\cos^{-1}(\frac{\sqrt{5}}{2\sqrt{2}})$, where we find ${\mathcal{S}}_{\overrightarrow{q}}=\frac{1}{\sqrt{10}}(1,2e^{i2uq+i\pi},
 2e^{i(8u+\frac{1}{2})q-i\pi},e^{i(4{u}-\frac{1}{2})q})\delta(\overrightarrow{q}-\overrightarrow{q_0})$ with $\overrightarrow{q}_0$ = $(\frac{2}{a_0}\cos^{-1}(\sqrt{\frac{5}{8}}),\frac{2}{a_0}\cos^{-1}(\sqrt{\frac{5}{8}}),0)$.  We can express the ``spins'' in real space as $\{s_i^{\delta},s_i^{\alpha},s_i^{\gamma},s_i^{\beta}\}$ = $\frac{1}{\sqrt{10}} e^{i2\cos^{-1}(\sqrt{\frac{5}{8}})(r_{ix}^{\delta}+r_{iy}^{\delta})}$ $\{1,-2,-2 e^{i4\cos^{-1}(\sqrt{\frac{5}{8}})}, $ $e^{i2\cos^{-1}(\sqrt{\frac{5}{8}})}\}$.  A phase change of $e^{i2\cos^{-1}(\sqrt{\frac{5}{8}})}$ (or a spin rotation of $75.5^0$) is gained by each site following a translation by one lattice constant  in the $\hat{x}$ or ${\hat{y}}$ direction of the cubic unit cell.  No change in the spin structure accompanies a translation by one lattice constant in the $\hat{z}$-direction.  This spin structure is plotted in Fig. \ref{figure8}(a).  

In this structure, the total spin on each triangle is 0, again satisfying Eq. \ref{equation6}.  However, ``spins'' on the $\delta$ and $\beta$ sites are seen to have only half the magnitude of those on the $\alpha$ and $\gamma$ sites (refer to Fig. \ref{figure1} for a definition of the $\{\delta,\alpha,\gamma,\beta\}$ sites).  
\subsubsection{(100)}

\begin{figure}
\includegraphics[scale=0.28]{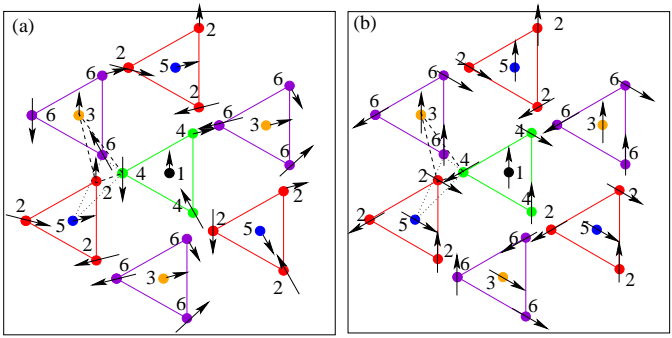}
\caption{\label{figure8} (Color online) Looking down the (111) axis of the trillium lattice, two degenerate ``spin'' configurations of the relaxed constraint. Numbers denote the layer number (or height) within the structure. Dashed and dotted lines label triangles out of the plane. (a) The $qq0$ state. (b) The $q00$ state. Note that the latter both forms a 120$^0$ state on each triangle and satisfies the hard spin constraint.     
}
\end{figure}

Along the (100) direction we find the energy minimum to occur at $q_0=\frac{2}{a_0}\cos^{-1}(\frac{1}{2})=\frac{2\pi}{3a_0}$, where ${\mathcal{S}}_{\overrightarrow{q}}=(\frac{1}{2},\frac{-1}{2},\frac{-ie^{i(2{u}-\frac{1}{4})q}}{2},\frac{ie^{i(2{u}-\frac{1}{4})q}}{2})\delta(\overrightarrow{q}-\overrightarrow{q_0})$.  Here $\overrightarrow{q_0}=(\frac{2\pi}{3a_0},0,0)$, implying that the real-space spin structure is $\{s_i^{\delta},s_i^{\alpha},s_i^{\gamma},s_i^{\beta}\}$ = $\frac{1}{2}e^{i\frac{2\pi}{3} r_{ix}^{\delta}}$ $\{1,e^{\frac{-i2\pi}{3}},1,e^{i\frac{2\pi}{3}}\}$.  A phase change of $e^{i\frac{2\pi}{3}}$ (or 120$^0$) is gained by each site following a translation by one lattice constant along the $\hat{x}$ direction of the cubic unit cell.  Translations by one lattice constant along the $\hat{y}$ and $\hat{z}$ directions leave the spin structure unchanged.
This spin structure is shown in Fig. \ref{figure8}(b).  This is a spin structure with 120$^0$ rotated spins of unit magnitude on every triangle.



\section{Mapping to a rigid rotor model}
\subsection{The hard spin contraint}
On frustrated lattices, subtle effects arise from constraints on the spins. On both the kagom{\'e} and corner-shared tetrahedral lattices the gournd state is known to be degenerate within both the (${\mathcal{N}}\rightarrow\infty$) mean field approximation and (${\mathcal{N}}=3$) classical Monte Carlo.  It is important to ask whether the ground state obtained by the mean field approximation on the trillium lattice can be reproduced by a method which satisfies the hard spin constraint.  Within the mean field approximation, the spin structures corresponding to the dispersion curves of Fig. \ref{figure4} satisfy the soft spin constraint.  We have seen from Section III D, that ``spin'' structures pertaining to a single ordering wavevector, $\overrightarrow{q}$, within the mean field, feature coplanar spins, and that the magnitude of these spins is generically found to vary from site to site within the unit cell.  However, the $(\frac{2\pi}{3a_0},0,0)$ spin structure of Fig. \ref{figure8} (b) was found to satisfy the hard spin constraint.  We see from the mean field that this structure, (and its equivalents: \{\{-$\frac{2\pi}{3a_0}$,0,0\},\{0,$\frac{2\pi}{3a_0}$,0\},\{0,$-\frac{2\pi}{3a_0}$,0\},\{0,0,$\frac{2\pi}{3a_0}$\},\{0,0,$-\frac{2\pi}{3a_0}$\}\}), is the only coplanar member of the ground state of the classical Heisenberg model satisfying a hard spin constaint.  To fully investigate the low energy properties of the Heisenberg model while strictly enforcing the hard spin constraint at each site, it is useful to consider the spins to be rigid rotors.  
\begin{figure}
\includegraphics[scale=0.4]{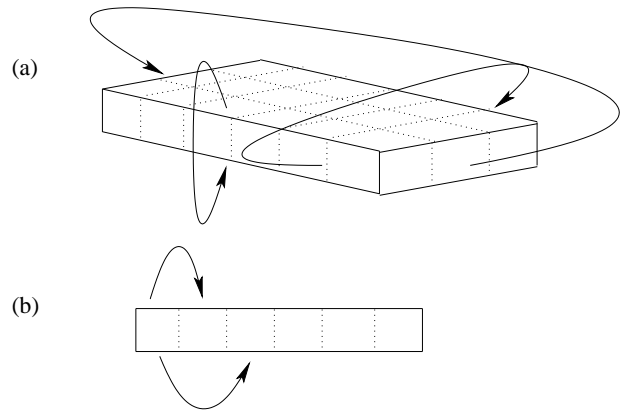}
\caption{\label{figure11}(a) To find the energy of a spin structure which is periodic along a particular direction, we design a finite size system satisfying periodic boundary conditions.  For a finite size spin lattice with $(L_x,L_y,L_z)$ unit cells in the $(\hat{x},\hat{y},\hat{z})$ directions, where $L_ia_0$ is the length of the system in the $\hat{i}$ direction, the spin structure repeats after $(L_x/n,L_y/m,L_z/l)$ unit cells where $\{n,m,l\}$ are integers greater than or equal to one.  Wavevectors with this periodic boundary condition satisfy ${\bf{q}}= (\frac{2\pi n}{L_x a_0},\frac{2\pi m}{L_y a_0},\frac{2\pi l}{L_z a_0})$. We then minimize the energy of randomly oriented spins on this lattice which chooses one (or more) value for each of $n$, $m$ and $l$.  As an example, the cubic lattice drawn above repeats after $(L,2L,1)$ unit cells and allows wavevectors of the form, ${\bf{q}}=(\frac{2\pi n}{La_0},\frac{2\pi m}{2La_0},0)$. (b) To only consider spin structures which repeat the same number of times in each (non-trivial) lattice direction $(n=m=l)$, we can write a one dimensional lattice with all unique unit cell positions and connections.  For example, for our cubic lattice which repeats after $(L,2L,1)$ unit cells, we can ask that the spin structure after a translation by one unit cell along the $\hat{x}$ direction is the same as the spin structure after a translation by two unit cells in the $\hat{y}$ direction.  Wavevectors, $\bf{q}$, with this additional periodic boundary condition satisfy $n=m$ as defined above, and correspond to a particular wavevector direction (in the example given here this is $(q,\frac{q}{2},0)$).  Minimization of such spin structures allows us to calculate the dispersion of the model with a hard spin constraint for several directions of ordering wavevectors.  }

\end{figure}
This allows us to numerically both: (a) investigate finite size lattices with periodic boundary conditions to determine whether the (100) ground state is truly unique or if non-coplanar spin structures (multiple wavevector) are able to satisfy the hard spin constraint while remaining in the ground state; and (b) investigate the low-lying energetic saddle-points for various wavevector directions.

\subsection{Method}

One way to strictly enforce the constraint of unit spins at each site is to represent the spins in terms of rigid rotors with fixed magnitude and two angles, $\theta$ and $\phi$, where $\overrightarrow{s}_i$=($\sin(\theta_i)\cos(\phi_i)$, $\sin(\theta_i)\sin(\phi_i)$,$\cos(\theta_i)$).  We minimize Eq. \ref{equation4} on finite size clusters in real space. With $N$ spins in a finite size lattice, the link matrix corresponding to the nearest neighbor couplings of the Heisenberg model is an $N\times N$ matrix. Such matrices can be simply written from the connections of Table \ref{Table II}. While this leads to large and rather unwieldy matrices, one can use periodic boundary conditions to reduce the large matrices to a smaller one-dimensional form, as shown in Fig. \ref{figure11}.   Varying the number of unit cells, $L$, in a manner consistent with the periodic boundary conditions allows one access to a finite but non-negligible number of low energy points in momentum space.  Accessible states have components of the form $q_i=\frac{2\pi n}{L_ia_0}$.   For each wavevector direction, $\hat{q}$, we minimize the energy with respect to the magnitude of $\overrightarrow{q}$.  We have chosen 7 unique directions for comparison with the mean field results.

\subsection{A unique $(\frac{2\pi}{3a_0},0,0)$ ground state}

Within the rotor model we have found that the ground state of the antiferromagnetic Heisenberg model on the trillium lattice is ordered along (100)\cite{period} with wavevector $q_0=\frac{2\pi}{3a_0}$ and features 120$^0$ rotated spins on every triangle.  The energy of this state is found to be $-3J$ as expected from Eq. \ref{equation1}.  The rotor model has the advantage that it gives us direct access to the spin structure, which is found to have the same form as shown in Fig. \ref{figure8} (b).  We note that this is the spin structure of the only member of the mean field ground state which satisfies the hard spin constraint.

\subsubsection{Energetic saddle-points: a sphere-like surface}

Any spin configuration within the rotor model automatically satisfies the hard spin constraint, so it is interesting to ask which features of the mean field approximation survive.  From the mean field, we found that restricting the angles to lie within a solid angle bounded by (100), (110) and (111), allowed access to all distinct energetics associated with the ground state.  It is thus natural to tile this area  and ask: (i) whether for each wavevector direction the energetic saddle-point occurs at approximately the same wavevector as given in Eq \ref{equation6}; (ii) about the relative magnitudes of the low lying excitation energies along the longitudinal (parallel to $\overrightarrow{q}$) and transverse (perpendicular to $\overrightarrow{q}$) directions; and (iii) whether there exists a simple analytic relationship between the mean field and lowest energy real spin structures.  

\begin{table}[hbtp]
\begin{tabular}{|l|l|l|l|l|}
\hline
dir&$E_{\text{min}}$&$|\overrightarrow{q}|_{\text{rotor}}^{(1)}$&$\overrightarrow{q}_{\text{rotor}}^{(2)}$&$|\overrightarrow{q}|_{\text{mf}}$\\
\hline
\hline
$qqq$&-2.6786$J$ &$\frac{\pi}{\sqrt{3}}\approx 1.814$&(0,0,0)&$\frac{\pi}{\sqrt{3}}$\\
$qq0$&-2.6666$J$&1.7198 &?&1.864\\
$q00$&-3$J$ &$\frac{2\pi}{3}\approx 2.09$&\text{none}&$\frac{2\pi}{3}$\\
$q\frac{q}{2}\frac{q}{2}$&-2.7632$J$ &1.776&\text{none}&1.862\\
$q\frac{q}{2}\frac{q}{4}$&-2.874$J$ &1.894&$(4\pi,2\pi,\pi)$&1.902\\
$q\frac{q}{2}$0&-2.9302$J$ &1.916&$(2\pi,\pi,0)$&1.926\\
$qq\frac{q}{2}$&-2.8525$J$ &1.767&$(2\pi,2\pi,\pi)$&1.835\\

\hline
\end{tabular}
\caption{\label{Table III} We see that the actual energetic minima along several wavevector directions continue to lie close to their mean field values, delineating a spheroid of saddle-points connected to the true mininum at $\frac{2\pi}{3a_0}$00. For simplicity we have set $a_0=1$.}
\end{table}

Table \ref{Table III} presents the results extracted in this fashion for  7 directions. (i) We see that the ground state  of $-3J$ is realized along (100) and that the extracted values of $|\overrightarrow{q}|^{(1)}_{rotor}$ closely follow those of the mean field, forming a sphere-like surface of saddle-points.  (We find that along (100) the spins are always coplanar so the mean field values are correct).  (ii) Remembering that the point ($\frac{\pi}{a_0}$,0,0) has the energy -2$\sqrt{2}J$, we see that spin excitations along (100) are quite soft and of a similar order of magnitude to those seen around the sphere-like surface of saddle-points. (iii) In the Section IV D we will present an analytic derivation of the minimal energy state along the $qqq$ symmetry axis of the crystal which corresponds to adding a $\overrightarrow{q}$=0 state to the mean field result.

\subsubsection{(100) saddle-points: a cross-check}

\begin{table}[hbtp]
\begin{tabular}{|l|l|l|l|}
\hline
lattice dimensions&$E_{min}^{rotor}$&$Min[\lambda_{q00}^{(1)},\lambda_{0q0}^{(1)},\lambda_{q00}^{(1)}]$&$q_{min}$\\
\hline
$2\times 2\times 2$&-2.8284271&$-2\sqrt{2}$&$\frac{\pi}{a_0}$\\
\hline
$1\times 4\times 4$&-2.9449468&-2.9449473&$\frac{\pi}{2a_0}$\\
\hline
$4\times 5 \times 5$&-2.9691749&-2.969174998&$\frac{4\pi}{5a_0}$\\
\hline
$1\times 10 \times 10$&-2.9915323&-2.9915325&$\frac{3\pi}{5a_0}$\\
\hline
\end{tabular}
\caption{\label{Table IV} Periodic clusters of up to 400 rigid three dimensional rotors of various shapes and sizes have been numerically minimized (see Fig. \ref{figure11} (a)) for the antiferromagnetic Heisenberg model. For representative clusters, we show: $L_x\times L_y \times L_z$, the number of 4-site unit cells considered along each axis $\{\hat{x},\hat{y},\hat{z}\}$ of the cubic lattice; $E_{min}^{rotor}$, the minimized rotor model energy; the lowest energy (100) state consistent with the boundary conditions; and $q_{min}$, the magnitude of the wavevector of this (100) state.  In all cases the minimized energy corresponds to the lowest energy  state which orders along (100) consistent with the boundary conditions.  Note that, in the last case, the boundary conditions allow the states with ${\bf{q}}=(0,\frac{\pm2\pi}{10a_0},\frac{\pm3\pi}{5a_0})$ which satisfy Eq. \ref{equation6} and were found to belong to the ground state in the mean field approximation, but the spin structure minimizes to one with ${\bf{q}}=(0,0,\frac{3\pi}{5a_0})$.  This shows that, with a hard spin constraint, low-lying excited states along (100) remain lower in energy than the saddle-points of the no longer degenerate spheroids. }
\end{table}

To check that the observation of order along (100) within the rotor model is not an artifact of the mapping procedure shown in Fig. \ref{figure11} (b), we have also considered finite size lattices as shown in Fig. \ref{figure11} (a).  The minimization of such lattices allows a spin structure to explore various multiple wavevector states that might be missed by the restricted geometry of Fig. \ref{figure11} (b).  In Table \ref{Table IV} we compare the energetics from minimizations of the three dimensional spin structures of various shapes and sizes.   An analytic expression for the dispersion along (100) which satisfies the hard spin constraint is given in Table \ref{Table V}.  This has been used to calculate the third column of Table \ref{Table IV} at the wavevector magnitude denoted $q_{min}$.  For all studied cases, the lowest energy spin structure of the rotor model is consistent with the lowest energy (100) state permitted by the boundary conditions. This shows that not only is the $(\frac{2\pi}{3a_0},0,0)$ ground state unique, but the longitudinal spin excitations along (100) are quite soft and dominate the low energy dispersion.

\begin{table}[hbtp]
\begin{tabular}{|l|l|l|}
\hline
a&Eigenvalue ($\lambda_{q00}^{(a)}$)& Eigenvector ($\psi_{q00}^{(a)}$)\\
\hline
1&$-2J(\cos(\frac{q}{2})+2\sin(\frac{q}{4}))$&$\frac{1}{2}(1,-1,ie^{-i(2{u}-\frac{1}{4})q},-ie^{-i(2{u}-\frac{1}{4})q})$\\
\hline
2&$2J(\cos(\frac{q}{2})-2\cos(\frac{q}{4}))$&$\frac{1}{2}(1,1,-e^{-i(2{u}-\frac{1}{4})q},-e^{-i(2{u}-\frac{1}{4})q})$\\
\hline
3&$-2J(\cos(\frac{q}{2})-2\sin(\frac{q}{4}))$&$\frac{1}{2}(1,-1,-ie^{-i(2{u}-\frac{1}{4})q},ie^{-i(2{u}-\frac{1}{4})q})$\\
\hline
4&$2J(\cos(\frac{q}{2})+2\cos(\frac{q}{4}))$&$\frac{1}{2}(1,1,e^{-i(2{u}-\frac{1}{4})q},e^{-i(2{u}-\frac{1}{4})q})$\\
\hline
\end{tabular}
\caption{\label{Table V} Analytic eigenvalues and eigenvectors of Eq. \ref{equation4} along (100).  These eigenvalues satisfy the hard spin constraint at every site and the lowest energy spin structures corresponding to the eigenvectors directly correspond to those found in the rotor model.  The longitudinal spin excitations from the ground state along this direction are at least as soft as excitations transverse to $(\frac{2\pi}{3a_0},0,0)$ as explained in Table \ref{Table IV}.  }
\end{table}

\subsection{Analytics: mean field to rotors}

Since real spins live in three-dimensional space, specifying one angle in principle only tells us information about the relative orientation of two components of the spin. We see from Fig. \ref{figure7} (a) that, with a soft spin constraint, the mean field ground state with (111) order had 120$^0$ rotated spins on triangles in the plane with normal vector (111) and no spin at the $\delta$ sites.  All other triangles featured two opposite spins.  It is instructive to add spin components normal to this plane to recover equal spins at each site.  That is, to obtain a spin structure with unit magnitude spins at each site, we perform a vector summation of the mean field spin structure of Fig. \ref{figure7} (a) and spin components orthogonal to this plane.  Clearly one must add a full spin to the $\delta$ site, while contributions to the $\{\alpha,\gamma,\beta\}$ sites must be added such that the spin components at each site square to one.  On the $\{\alpha,\gamma,\beta\}$ sites, we assign an amplitude $\sin(\theta)$ to spins of the mean field structure and an amplitude $\cos(\theta)$ to the spin component perpendicular to the plane.  We refer to this perpendicular component of the spins as a $q=0$ component because, despite structure within the 4-site cubic unit cell, it is unchanged from one unit cell to another.    
\begin{figure}
\includegraphics[scale=0.35]{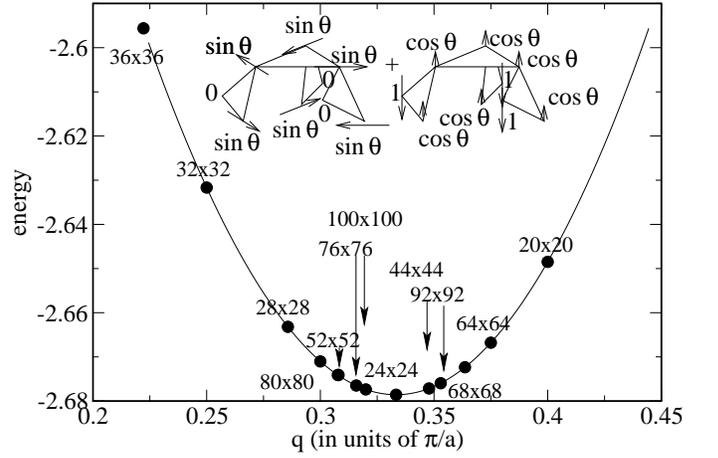}
\caption{\label{figure16} (inset) An illustration of the spin components of the noncoplanar spin structure at $(\frac{\pi}{3a_0},\frac{\pi}{3a_0},\frac{\pi}{3a_0})$ on the four unique triangles of the trillium lattice. Analytic results have been derived by the addition of a perpendicular $\bf{q}$=$\bf{0}$ spin component (see main text) to the mean field results of Fig. \ref{figure7} (a), to enforce the hard spin constraint. In the main figure we present a comparison of the rotor model results (points) with analytic results of Eq. \ref{equation29} obtained by minimization of $\theta$ as explained in the text. The $N\times N$ link matrix written above each point shows the number of spin sites of each one dimensional lattice under consideration.  The corresponding wavevector along (111) can be extracted as $q_{rotor}^{(1)}=\frac{2\pi n}{(N/4)a_0}$, where $n$ is an integer.
 }
\end{figure}
This is shown pictorially in the inset to Fig. \ref{figure16} on the four different magnetic triangles of the trillium lattice.  Then we can write the total energy per unit spin as a function of the angle $\theta$ as,
\begin{equation}
\epsilon= \frac{-9\sin^2(\theta) + 2(6 \cos^2(\theta)+ 6 \cos(\theta))}{4}\label{equation27}
\end{equation} 
and to find a true minimum of this function, we simply solve for the derivative,\begin{equation}
\frac{d\epsilon}{d\theta} = -\frac{21\sin(\theta)\cos(\theta)+6\sin(\theta)}{2}=0\label{equation28}
\end{equation}
which yields $\theta=\cos^{-1}(\frac{-2}{7})$.  Substitution of $\theta$ in Eq. \ref{equation27} yields an energy of $-2.6786J$.  This is the same as the lowest energy solution found by the rotor model along (111), and is the true minimal energy state with this wavevector direction.  

Within the mean field approximation, the lowest energy solution of Eq. \ref{equation4} can be shown to be $E_{mf}(qqq) = -2J(\cos^2(\frac{q}{2})+\sqrt{3 \cos^2(\frac{q}{2})\sin^2(\frac{q}{2})})$.  The spin structure remains as presented in Section III D, with the exception of the relative angle change of the spin from one unit cell to the next; for generic values of $q$ the triangles which include the $\delta$ site no longer feature parallel spins.
As we have analytic expressions for our eigenvalues, it is interesting to write the general expression in the (111) direction following the same steps, 
\begin{eqnarray}
\epsilon&=&3(1-(\frac{1}{2(\cos(\frac{q}{2})\cos(\frac{q}{2}+\frac{2\pi}{3})-1)})^2)\cos(\frac{q}{2})\times\nonumber\\&&\cos(\frac{q}{2}+\frac{2\pi}{3})+3(\frac{1}{(2\cos(\frac{q}{2})\cos(\frac{q}{2}+\frac{2\pi}{3})-1)})^2\nonumber\\&& + \frac{3}{2(\cos(\frac{q}{2})\cos(\frac{q}{2}+\frac{2\pi}{3})-1)},\label{equation29}
\end{eqnarray}
which is plotted in Fig. \ref{figure16} and agrees very well with the rotor model.

Note that in principle this must be done also within the mean field solution of Reimers' paper{\cite{reimers}}, in such a way as to satisfy the hard spin constraint on the pyrochlore lattice. That is, an analogous mean field solution is found along $(qqq)$ at $q=\frac{\pi}{3a_0}$ in the pyrochlore case, but the high degeneracy of the lattice allows the addition of a $q=0$ (ie. the same in each unit cell) components to satisfy the hard spin constraint.  Relative to mean field calculations, Monte Carlo calculations (which satisfy a hard spin constraint) on the pyrochlore lattice would be expected to differ by a small factor.  That this has not previously been appreciated is likely due to the high degeneracy of the pyrochlore lattice--it is not nearly as important an effect there as it is here.

\section{Concluding remarks}
Motivated by recent neutron scattering and spin susceptibility measurements of MnSi, we have considered the role of antiferromagnetic interactions and their possible relevance to a partially ordered state in this system.  To address this issue, we have studied a Heisenberg model on the trillium lattice, a new three dimensional corner-shared triangle lattice formed by the Mn atoms of MnSi.  The trillium lattice is also a sublattice of other systems including the CO (B21), NH$_3$ (D1), NiSSb (F0$_1$, Ullmanite) and FeSi (B20) structures.

The Heisenberg model is an oversimplified model for an itinerant system such as MnSi.  However, this model has been used to reproduce the measured spin correlations in other itinerant systems including (Y$_{0.97}$Sc$_{0.03}$)Mn$_2$, LiV$_2$O$_4$ and $\beta$-Mn, where it has been considered the first step in an understanding of the magnetic correlations present.  We found that a certain degree of geometric frustration does exist on the trillium lattice with a na{\"i}ve estimation of the number of degrees of freedom available to the ground state less than the kagom{\'e} and pyrochlore lattices, but on a par with the $\beta$-Mn lattice and greater than the hexagonal lattice.

We have treated the classical antiferromagnetic Heisenberg model on the trillium lattice within the mean field approximation and by mapping the model to a lattice of rigid rotors. 
  Within mean field theory on the trillium lattice, this model was found to have a degenerate ground state with wavevectors found to lie around a sphere-like surface given by Eq. \ref{equation6}.  At small but finite temperatures, the neutron scattering structure factor would then be expected to feature weight predominantly confined to lie on this surface.  Along this sphere-like surface, a disperse weight modulation would arise solely from geometrical factors and be roughly independent of temperature as found of the ``bow-ties'' of the corner-shared tetrahedral lattice.  Measurements normal to these sphere-like shells would show resolution limited sharp peaks as the temperature decreased as shown in Fig. \ref{figure6}.  These features bear a strong qualitative resemblance to neutron scattering measurements near the critical pressure of the itinerant helimagnet, MnSi, although notable differences are present.  Chief among these are the radius of the observed sphere-like shape, which is smaller by a factor of 10, and the detailed nature of the pattern seen.  More precisely, the measurements{\cite{pflei}} appear to have been made with respect to the lattice Bragg peaks at ($\frac{2\pi}{a_0}$,$\frac{2\pi}{a_0}$,0). About this point there is no reason to expect an inversion symmetry of the structure factor and the C$_3$ symmetry of the lattice which is responsible for the vanishing of weight along (111) shown in Fig. \ref{figure5} would not be present.

By mapping the classical antiferromagnetic Heisenberg model to a rigid rotor description with a spin of unit magnitude at each site of the trillium lattice, we have found an ordered ground state at wavevector{\cite{period}} ($\frac{2\pi}{3a_0},0,0$) featuring 120$^0$ rotated spins on each triangle.  The rotor model is a real space construction which imposes periodic boundary conditions on finite size clusters of three dimensional vectors representing the spin at each site of the lattice.  For several small cluster sizes (see Table \ref{Table IV}), we have found the minimal energy state to be consistent with order along (100) as analytically described in Table \ref{Table V}, indicating that low-lying longitudinal fluctuations from the ground state are quite soft.  We have designed an extension of the rotor model in such a way as to allow only one particular wavevector direction at a time.  That is, rather than studying an $M\times N\times P$ lattice, where $\{M,N,P\}$ are integers counting the number of cubic unit cells in each direction, we have used symmetry to map the lattice to a one dimensional lattice of unit cells.  For example, to investigate spiral order along (111), unit cells at $\{(1,0,0),(0,1,0),(0,0,1)\}$ are all identical and can be replaced in the mapping by a single unit cell.  Despite the strength of (100) order, this allows us access to low-lying energetic saddle-points.  In Table \ref{Table III} we have shown that the wavevector associated to these saddle-points continues to lie close to solutions of Eq. \ref{equation6}.  An analytic derivation of the wavevector and spin structure at the saddle-point along (111) has been developed in Section III D which illustrates the relationship between mean field and rotor model results.

The discrepancy between the results obtained via the mean field approximation and those obtained by the mapping to the rotor model arises from the use of soft and hard spin constraints in the respective approximations.  The soft spin constraint allows the magnitude of the spin to vary within the unit cell as illustrated by the ``spin'' structures of Section III D.  For itinerant magnetic systems, it is not unusual for the magnetization density to vary spatially within an effective magnetic model.  The hard spin constraint of the rotor model sets the magnitude of the spin at each site to be 1.

It is interesting to note that problems associated with the soft spin constraint of the mean field approximation arise in earlier treatments of frustrated lattices, yet do not seem to have been commented on.  This is due, in part, to the extensive ground state degeneracy of these systems.  For example, on the corner-shared tetrahedral lattice (a sublattice of the pyrochlore structure), the mean field approximation yields{\cite{reimers}} a doubly degenerate flat band with a three band degeneracy at $q=0$.  Even though the coplanar mean field ground state spin structure along (111) does not satisfy a unit spin magnitude ($s_i^2=1$) at one of the four sites on each tetrahedron, it is possible to add spin components (with $q=0$) to every site to recover this constraint.  In this way, one finds a spin structure which belongs to the ground state and has components of its spins which rotate along (111).  Calculations of the neutron scattering structure factor{\cite{isakov}} within the mean field approximation make the implicit assumption that the entire spin rotates.  That along certain wavevector directions only a component of the spin is rotating must imply that the detailed nature of the relative neutron scattering weight is not fully captured within the mean field approximation, despite the remarkable qualitative agreement seen with Monte Carlo results.  Discrepancies between the soft and hard spin constraints are simply more clearly evident on the trillium lattice, as satisfying the hard spin constraint lifts the degeneracy of the ground state.

The advantage of the mean field approximation is that it allows one to access all wavevectors and easily generalizes to finite temperatures.  Indeed, at temperatures $T\ge0.4J$ one sees, from the rotor model (see Table \ref{Table III}), that all wavevector orientations should become accessible to the system.  One might hope at such temperatures to recapture many of the qualitative features of the mean field description.  A finite temperature comparison between the mean field approximation and Monte Carlo results will be presented in the near future{\cite{Sergeius}}.  Future work{\cite{usto}} will treat the extended Heisenberg model which is expected to exhibit a rich phase diagram, possibly relevant to currently available magnetic monosilicides.

It is worthwhile to mention the possibility of frustrated ferromagnetically coupled spin systems forming on this lattice.  Indeed, on the pyrochlore lattice there are arguably more candidates for spin-ice physics than for degenerate antiferromagnetic ground states.  As we will show in future work\cite{tbpubl}, if Ising spins are restricted to lie along the local (111) axes of the crystal (this is the direction of the closest non-magnetic atoms in some of the Ullmanites and may arise due to crystal/ligand field splittings), then antiferromagnetically coupled spins will be expected to order while ferromagnetically coupled spins will be extensively degenerate possessing a generalized version of the two-spins out, two spins in rule of the corner-shared tetrahedral lattice.

\vskip1pc
{\centering
{\bf{ACKNOWLEDGMENTS}}\\}
\vskip1pc

J.H. would like to thank Krishnendu Sengupta, Sergei Isakov and Benedikt Binz for many useful discussions.  We are grateful to B. Canals and C. Lacroix for pointing us towards Ref. [\cite{naka}] and [\cite{canlac}], D. van der Marel for providing us with a copy of his results prior to their publication, and D. Khomskii for noticing that the next smallest loop after triangles is 5 sided.  This work was supported by an NSERC PDF, an NSERC CRC, an Alfred P. Sloan Fellowship, NSERC and the CIAR.



\begin{thebibliography}{00}
\bibitem{trill} About any site of the lattice, three equilateral triangles join in a corner-shared manner reminiscent of the trillium.  Both the foliage and flowers of the trillium feature three parts which join at the stem.  This naming allows us to distinguish this unstudied lattice from the distorted windmill structure named in Ref. {\cite{naka}} and studied in Ref. {\cite{canlac}}.
\bibitem{explan} Although the ordered magnetic moment is found to be quite small{\cite{thess,vanderMarel}} ($\approx 0.4 \mu_B/Mn$), above T$_c$, the Curie-Weiss slopes are consistent with a much larger Curie-Weiss moment ($\approx 2.5\mu_B/Mn$){\cite{vanderMarel,vanderMarel2}} which seems consistent with spin-1 local moments at each site at higher temperatures.
\bibitem{equivalent} Note that the four sites of the corner-shared tetrahedral lattice are ``equivalent'' after a rotation of their symmetry axes.  In exactly the same manner, the four sites of the trillium lattice are ``equivalent''.
\bibitem{period} This state is equivalent to (-$\frac{2\pi}{3},0,0$), (0,$\pm\frac{2\pi}{3},0$), (0,0,$\pm\frac{2\pi}{3}$), where the latter four states arise due to the $C_3$ symmetry of the lattice and the signs reflect the chirality of the spin structure.
\end{thebibliography}
\end{document}